# Performance Evaluation of Unicast and Broadcast Mobile Ad-hoc Networks Routing Protocols

Sumon Kumar Debnath[1], Foez Ahmed[2], and Nayeema Islam[3]

[1]Dept. of Computer Science and Telecommunication Engineering, Noakhali Science and Technology University, Bangladesh
[2]Dept. of Computer Networks Engineering, College of Computer Science, King Khalid University, Kingdom of Saudi Arabia
[3]Department of Information & Communication Engineering, Rajshahi University, Rajshahi- 6205, Bangladesh

*Abstract*— Efficient routing mechanism is a challenging issue for group oriented computing in Mobile Ad Hoc Networks (MANETs). The ability of MANETs to support adequate Quality of Service (QoS) for group communication is limited by the ability of the underlying ad-hoc routing protocols to provide consistent behavior despite the dynamic properties of mobile computing devices. In MANET QoS requirements can be quantified in terms of Packet Delivery Ratio (PDR), Data Latency, Packet Loss Probability, Routing Overhead, Medium Access Control (MAC) Overhead and Data Throughput etc. This paper presents an in-depth study of one-to-many and many-to-many communications in MANETs and provides a comparative performance evaluation of unicast and broadcast routing protocols. Dynamic Source Routing protocol (DSR) is used as unicast protocol and BCAST is used to represent broadcast protocol. The performance differentials are analyzed using ns-2 network simulator varying multicast group size (number of data senders and data receivers). Both protocols are simulated with identical traffic loads and mobility models. Simulation result shows that BCAST performs better than DSR in most cases.

*Keywords-MANETs, DSR, BCAST, Unicast, Broadcast, Random way point model*

I. INTRODUCTION

A mobile ad hoc network is a self-organizing network comprising wireless mobile nodes that move around arbitrarily and can able to communicate among themselves using wireless radios, without the aid of any preexisting infrastructure [1]. Each participating mobile node can act as sender, receiver and even as a router at the same time and able to build, operate and maintain these networks [2].

Due to limited radio coverage of these wireless devices efficient support of group oriented communication is extremely critical in most MANET applications. In MANET group communications issues differ from those in wired environments for the following reasons: The wireless medium has variable and unpredictable characteristics. The signal strength and propagation fluctuate with respect to time and environment resulting disconnection of the network at any time even during the data transmission period [3]. The strength of the received signal depends on the power of the transmitted signal, the antenna gain at the sender and receiver, the distance between two mobile nodes, the obstacles between them, and the number of different propagation paths the signals travel due to reflection. Further node mobility also creates a continuously changing communication topology in which existing routing paths break and new ones form dynamically. Since MANETs have limited channel bandwidth availability and low battery power, their algorithms and protocols must conserve both bandwidth and energy [3]. Wireless devices usually use computing components such as processors, memory, and I/O devices, which have low capacity and limited processing power. Thus their communication protocols should have lightweight computational and information storage capability fulfilling some key features like robustness, simplicity and energy conserving. For this reason, several prominent unicast, multicast and broadcast protocols deployed in static wired networks that can not perform well in ad-hoc networks [3, 4].

In-group oriented communication system, routing protocols can be classified into three main categories [5, 6, 7] based on the number of senders and receivers in MANETs. Unicast communication is the point-to-point transmission with one sender and one receiver. While unicasting is a simple mechanism for one-to-one communication, for one-to-many or many-to-many dissemination it brings the network to its knees due to huge bandwidth demands [8]. This can also introduce significant traffic overhead, sender and router processing, power consumption, high packet latency and poor throughput in the network. To minimize these overhead for one-to-many or many-to-many communication multicast and broadcast Ad-Hoc routing protocols play an important role. Multicast communications are both one-to-many and many-to-many traffic pattern [9] i.e. to transmit a single message to a select group of recipients where as in broadcast routing communications is one-to-all traffic pattern. It is a basic mode of operation in wireless medium that provides important control and route establishment functionality for a number of unicast and multicast protocols. When designing broadcast protocols for ad hoc networks, developers seek to reduce the overhead such as collision and retransmission or redundant retransmission, while reaching all the network's nodes. In practice, the scope of the broadcast is limited to a broadcast domain. Broadcasting is largely confined to local area network (LAN) technologies, most notably Ethernet and Token Ring, where the performance impact of broadcasting is not as large as it would be in a wide area network. Because Broadcasting is used to carry huge amount of traffic and requires more bandwidth, neither X.25 nor frame relay supply a broadcast capability, nor Internet explicitly support broadcasting at the global level [10].

This paper compares two Ad Hoc routing protocols: unicast reactive DSR and BCAST protocol over group oriented communication system. Performance comparisons are based on Shadowing path loss model and Random way point mobility model. The simulation of two routing protocols focuses on their differences in their dynamic behaviors that can lead to performance differences.

## II. UNICAST AND BROADCAST ROUTING PROTOCOLS

### A. Dynamic Source Routing(DSR) Protocol

DSR [11, 12] is an on-demand unicast reactive source-routed routing protocol. This means the source node determines the complete sequence of route information between source and destination and explicitly lists each hop of the path in the packet's header. Route is determined dynamically without any prior configuration necessary. The intermediate nodes do not require huge memory resources because they do not need to maintain consistent global routing information in order to route the packets that they forward. The basic operation of DSR contains two phases: route discovery and route maintenance. Route Discovery mechanism is used to find a source route to destination only when source attempts to send a packet to destination and does not already knows a route. To reduce the cost of Route Discovery, each node maintains a Route Cache of source routes it has learned or overheard. Route Maintenance is the mechanism used to detect if the network topology has changed such that it can no longer use its route to the destination because some of the nodes listed on the route have moved out of range of each other.

### B. Broadcast Routing ( BCAST) Protocol

BCAST is an optimized scalable broadcast routing protocol [13]. It keeps track of one-hop and two-hop neighbor knowledge information that are exchanged by periodic "Hello" messages. Each "Hello" message contains the node's IP address and list of known neighbors. When a node receives a "Hello" packet from all its neighbors, it has two-hop topology information i.e. only packets that would reach additional neighbors are re-broadcast. For example if a node, *B* receives a broadcast packet from another node *A*, it knows all neighbors of *A*. If *B* has neighbors not covered by *A*, it issues the broadcast packet with a random delay. During this delay, if *B* receives another copy of this broadcast from *C*, it can check whether its own broadcast will still reach new neighbors. If this is no longer the case, it will drop the packet. Otherwise, the process continues until *B*'s timer goes off and *B* itself rebroadcasts the packet.

The determination of Random delay time is very critical. To solve this problem a dynamic strategy is suggested in literature [13]. Each node searches its neighbor table for the maximum neighbor degree of any neighbor node, say *MAX*. If its own node degree is N, it calculates the random delay as *MAX/N*. Every node also buffers the most recent *X* packets. *X* can be any arbitrary integer number. To keep the memory requirement at each node low; set *X* to a small number. This mechanism improves the packet delivery ratio in BCAST.

When a node receives a packet with sequence number *N* from source node *A*, it checks whether it also received packet *N*-1 from the same source. If not, it issues a one-hop broadcast to the neighbors, asking for retransmission of this packet by sending Negative Acknowledgement, NACK(N-1, A) message. Each neighbor, upon receiving the NACK packet, will check its local buffer and if they have this packet buffered, will schedule a retransmission. To reduce collisions, the NACKs and the packet retransmissions are jittered randomly by few milliseconds.

## III. SIMULATION MODEL

This section describes the simulation tools and parameters chosen to simulate the routing protocols

### A. Simulation Environment

Network Simulator NS-2 [14, 15] is chosen to compare DSR and BCAST routing protocols. NS-2 is discrete event packet-level simulators with CMU's Monarch group's mobility extensions. It includes implementations of models of signal strength, radio propagation, wireless medium contention, capture effect, and node mobility. A simulation model with MAC and physical models are used to study the interlayer interaction and their performance. An unslotted carrier sense multiple access (CSMA) technique with collision avoidance (CSMA/CA) is used to transmit the data packets. In this experiment, the Distributed Coordination Function (DCF) of IEEE 802.11 for wireless LAN is used as MAC layer. The simulated radio interface model is the Lucent WaveLAN. WaveLAN is modeled as shared-media radio with channel capacity of 2Mbits/sec and transmission range of 250m and the carrier sensing range is 471.5m. All protocols use an interface queue (IFQ) of 50 packets. The IFQ is a FIFO priority queue where routing packets gets higher priority than data packets. All MAC and Network layer operations of the wireless network interfaces are logged in trace files.

### B. Radio Propagation Model

The shadowing path loss model [15] is used in this simulation study. It attempts more realistic situation than free space and two-ray path loss models. It takes into account multi-path propagation effects. Both free space and two-ray models predict the mean received signal strength as a deterministic function of distance and consequently represent communication radius as an ideal circle. But in realistic environment, when the fading effects are considered it can be seen that, the received power at a certain distance is a random variable. Hence shadowing model is widely used in real environment.

The available parameters that are used in our simulation code are shown in table I.

TABLE I. PARAMETERS USED IN SIMULATION

| Parameters | Value | Comment |
|---|---|---|
| Transmission Range | 250m | Fixed (Considered) |
| Frequency | $914 \times 10^6 Hz$ | Fixed (Considered) |
| Path Loss Exponent | 2.0 | Fixed (Considered) |
| Standard Deviation | 4.0 | Fixed (Considered) |
| Reference Distance | 1.0m | Fixed (Considered) |
| CPThreshold | 10.0 Watt | Fixed (Considered) |
| RXThreshold | $6.76252 \times 10^{-10}$ | Calculated |
| CSThreshold | $2.88759 \times 10^{-11}$ | Calculated (RXThreshold*0.0427) |
| Power (Pt) | 0.28183815 Watt | Fixed (Considered) |
| System Loss | 1.0 | Fixed (Considered) |

*C. Traffic and Mobility Model*

In this simulation Continuous bit rate (CBR) traffic sources are used. The source-destination pairs are spread randomly over the network. Only 512- byte data packets are used. The number of source destination pairs and the packet-sending rate in each pair is varied to change the offered load in the network.

A mobility model accurately represents the movement of mobile nodes in MANET. Random waypoint mobility model [16] *(*RWM) is used in this simulation study. The model includes networks with 50 mobile nodes placed on a site with dimensions 1500×300 meters. Each packet starts its journey from a random location to a random destination with a randomly chosen speed (uniformly distributed between 0–20 m/s). Once the destination is reached, another random destination is targeted after a pause time and then repeats the process. The pause time, which affects the relative speeds of the mobiles, is also varied. Five randomly generated scenarios are run for each parameter combination, and each point in the graphs is the average results of these five scenarios. Identical mobility and traffic scenarios are also used across protocols to gather fair results. In RWM model, *Pause Time* and *Max Speed* of a mobile are the two key parameters that determine the mobility behavior of nodes. If the node movement is small and the *Pause Time* is long, the topology of Ad Hoc network becomes relatively stable. On the other hand, if the node moves fast and the pause time is small; the topology is expected to be highly dynamic.

IV. PERFORMANCES METRICES

The performance of DSR and BCAST protocols are compared using the following important Quality of Service (QoS) metrics

**Packet Delivery Ratio (PDR):** The ratio of the number of packets received by the CBR sinks at the final destination to those generated by the CBR sources.

**Packet Latency:** This includes all possible delays caused by buffering during route discovery, queuing delay at the interface queue, retransmission delays at the MAC, propagation and transfer times [17]. The lower the packet latency the better the application performance as the average end-to-end delay is small.

**Normalized routing Load** (NRL): The ratio of the number of routing packets sent to the number of data packets received. Each hop-wise transmission of these packets is counted as one transmission [18].

**Normalized MAC Load** (NML): The number of routing, Address Resolution Protocol (ARP), and control packets (e.g., RTS, CTS and ACK) transmitted by the MAC layer, including IP/MAC headers for each delivered data packet [18]. It considers both routing overhead and the MAC control overhead. This metric also accounts for transmission at each hop.

**Throughput**: The ratio of the total data received by the end user and the connection time [19]. A higher throughput directly impacts the user's perception of the QoS.

V. SIMULATION RESULTS AND DISCUSSIONS

The results of this simulation study are separately considered into two sections.
- Varying Number of data Senders:
- Varying Number of data Receivers

*A. Effect of number of senders on QoS metrics*

To investigate the effect of number of senders on the performance of DSR and BCAST, the data send rate and number of data receivers is kept constant at 2 packets/sec. and 20 respectively. The numbers of data senders are increased from 1 to 10 and several QoS metrics are measured and plotted into logarithmic graphs. For the fairness of protocols comparison and network performance, each ad hoc routing protocol is run over the same set of scenarios. Table II. Shows Simulation parameters for the different Senders Scenarios

TABLE II. SIMULATION PARAMETERS FOR THE DIFFERENT SENDERS SCENARIOS

| Parameter | Value |
| --- | --- |
| Number of senders (variable) | 1, 2, 5, 7 and 10 |
| Number of receivers (keep constant) | 20 |
| Pause Time | 0 m/s |
| Max. Speed | 20 m/s |
| Antenna Range | 250 m |
| CBR Rate | 2 packets/sec. |
| Simulation Time | 200 s |

The PDR and data packet latency simulation results as a function of number of Senders are given in fig 1.

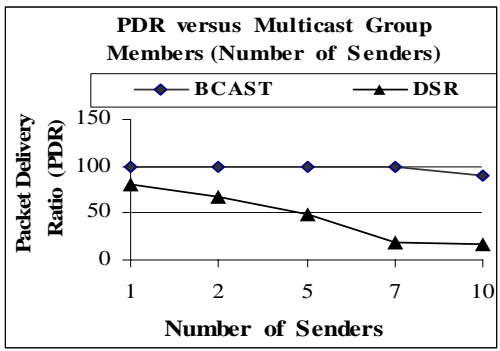

(a)

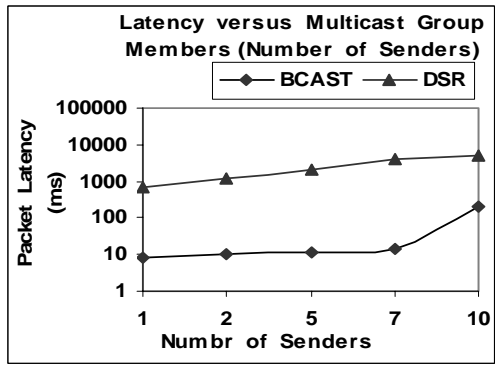

(b)

Figure 1. PDR (a) and Latency (b) results as a function of Number of Senders.

From fig 1(a) it is observed that with increasing number of senders, the PDR of BCAST protocol is higher and relatively consistent as compare to DSR routing protocol. A relatively high PDR is a desirable property for the scalability of Ad Hoc routing protocols. Unicast DSR shows lower PDR than BCAST for one-to-many and many-to-many communication. This is due to the fact that with relatively many senders (two or more) the traffic sources are more spread throughout the MANET and hence the overall performances of on-demand unicast DSR deteriorate rapidly. For example in DSR with five senders and 30 receivers, each sender needs to generate and maintain 30 simultaneous unicast connections to connect 30 multicast group members. Hence overall 150 unicast connections are required. Each data source also requires generating 30 data packets. This provides an inefficient use of wireless medium and causes congestion in the network. Since for many senders the PDR value drops drastically, the performance of DSR protocol is not very attractive for many-to-many application and it also increases the packet transmission cost. BCAST shows higher packet delivery ratio for most scenarios. Fig 1(a) also shows that for five senders, the PDR of BCAST and DSR are 99.45% and 47.82% respectively. This is because that BCAST has less redundancy and dynamically selects only a subset of nodes to re-broadcast a packet. It keeps 2-hop neighbor topology information and each node also buffers most recent few packets. A NACK based retransmission scheme of BCAST protocol further increase PDR.

From fig 1(b) it is observed that the average packet latency increases with increasing number of senders. BCAST protocol performs better than DSR in this case. For example for seven senders, the packet latency of BCAST and DSR are 14.43 ms and 3.86 $\mu s$ respectively. This is due to the fact that DSR maintains unicast connections. As the number of sender's increases, it requires to generate more packets in order to reach the group members, more routing packets causes delay in the interface queue before reaching the intended destination. This causes more packet delay with increasing the number of senders. Since BCAST maintains broadcast connections and keep two hops topology information, the average packet delay is significantly lower than DSR. Lower packet latency is the desirable property for real-time applications because these applications can tolerate loss but very sensitive to delay. Hence BCAST is more effective for real time applications.

The NRL and NML simulation results as a function of number of Senders are given in fig 2.

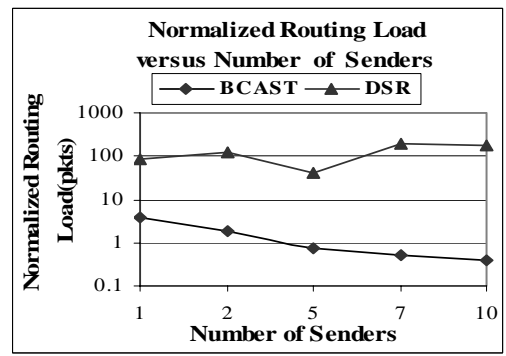

(a)

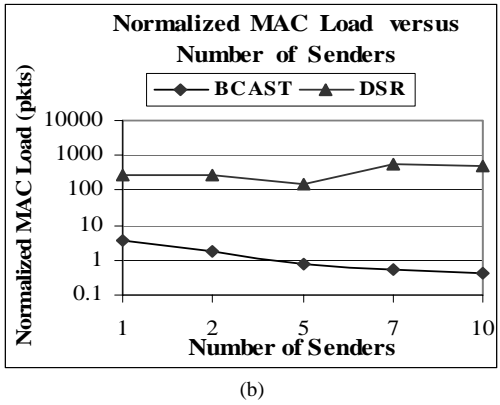

Figure 2. NRL (a) and NML (b) results as a function of Number of Senders

From fig 2(a) it can be observed that the NRL of DSR is higher than BCAST and hence provides poor performance. This is because that with increasing the number of senders DSR requires more routing packets to maintain unicast connections among group members. For example for five senders, NRL of DSR and BCAST are 44.23 packets and 0.735 packets respectively. From fig 2(b), it is also observed that in DSR normalized MAC load is also extremely high than BCAST protocols. In this case almost all MAC transmissions are unicast, a high fraction these transmitted packets are MAC layer control packets (RTS, CTS and ACK). Due to unicast nature, as the number of senders increase DSR requires to generate more MAC control packets. BCAST gives better performance in this case. Since in BCAST all MAC transmissions are multicast, it generates only fraction of MAC layer control packets than DSR. This effect results lower transmission collision and offers high packet delivery guarantee. For example for 10 senders the NML of BCAST and DSR protocols are 0.41 and 511.03 packets respectively.

The data throughput results as a function of number of Senders are given in fig 3.

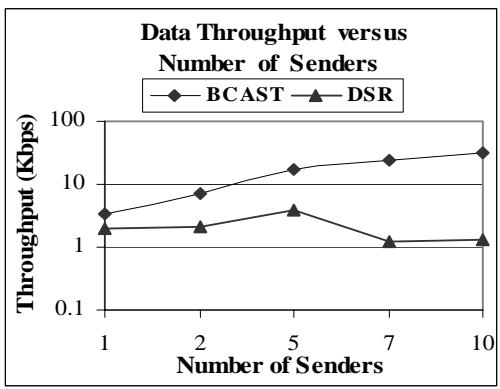

Figure 3. Data throughput results as a function of Number of Senders.

This throughput result is consistent with the control overhead (NRL and NML) shown in fig 3. This is because that the decay of overhead is directly related to the throughput growth. From fig 3 it can also be observed that with increasing the number of senders, the data throughput increases and the throughput of DSR does not increases too drastically and provides poor performance. For few senders (one to five) DSR throughput increases but lower than BCAST protocols. With more senders the DSR throughput highly degrades. Since DSR packet drop probability increases with increasing the number of senders, it affects the DSR throughput. It can be observed that as the number of senders increase, the throughput of BCAST increases and the maximum throughput are achieved for BCAST protocol. For example at five senders, fig 3 mentions that the data throughputs of BCAST and DSR protocols are 17.54 kbps and 3.91 kbps respectively.

### B. Effect of number of receivers on QoS metrics

In this case, to evaluate the performance of DSR and BCAST, the data send rate and number of data senders are kept constant at 2packets/sec. and 05 respectively. The numbers of multicast receivers are varied and QoS metrics are measured.

The simulation parameters considered for performance evaluation are provided in table III.

TABLE III. SIMULATION PARAMETERS FOR THE DIFFERENT RECEIVERS SCENARIOS

| Parameter | Value |
| --- | --- |
| Number of senders (keep constant) | 05 |
| Number of receivers (variable) | 10, 20, 30, 40 and 50 |
| Pause Time | 0 m/s |
| Max. Speed | 20 m/s |
| Antenna Range | 250 m |
| CBR Rate | 2 packets/sec. |
| Simulation Time | 200 s |

The PDR, average packet latency and throughput results as a function of number of receives is shown in fig 4.

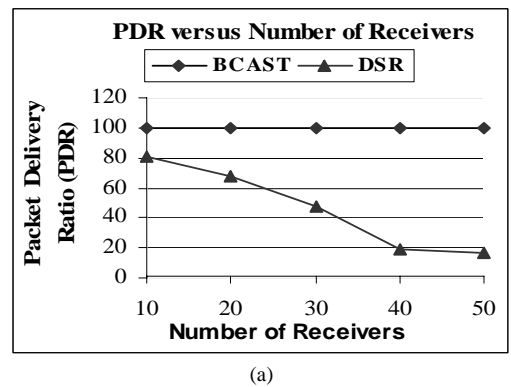

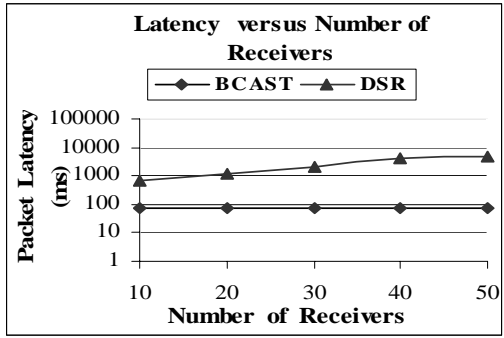

(b)

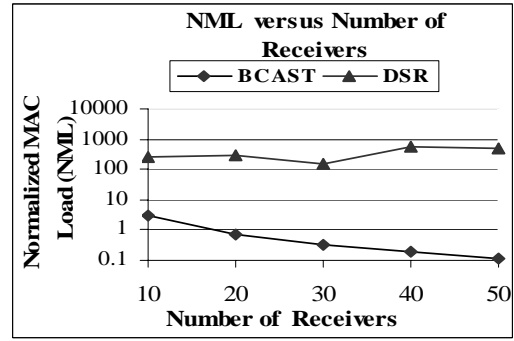

(b).

Figure 5. .NRL (a) and NML (b) results as a function of versus Number of Receivers

From fig 5(a) and fig 5(b), it is observed that the normalized routing and MAC load of DSR and BCAST protocols are also similar as in the case of number of senders.

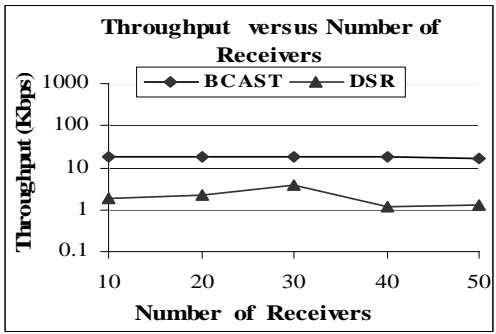

(c)

Figure 4. .PDR (a), Latency (b) and Data throughput (c) results as a function of Number of Receivers.

From fig 4(a) and fig 4(b), it is observed that, DSR gives similar PDR and Latency results as in the case of number of senders. BCAST protocol also shows better performance and these performances are independent of the number of multicast receivers. From fig 4(c), it is observed that, the throughput performance of BCAST is independent of the number of multicast receivers. BCAST protocol shows better performance in this case. In DSR when the number of multicast receivers below 30, DSR throughput performance increases with increasing the number of receivers. After this the throughput deteriorates again.

The NRL and NML simulation results as a function of number of multicast receivers are given in fig 5.

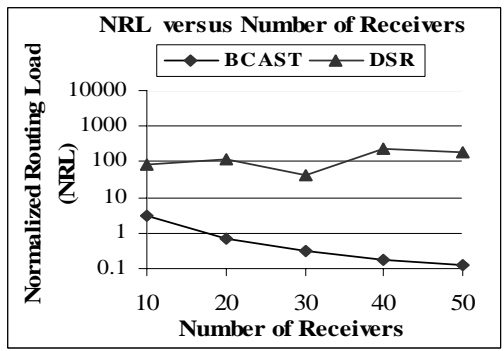

(a)

## VI. CONCLUSIONS

There are number of alternatives when delivering data from one or more senders to a group of receivers such as setting up dedicated unicast connections from each sender to each receiver, employing a unicast, multicast protocol and broadcasting packet to every node. This paper compares the performance of BCAST and DSR routing protocols over group communication in MANETs. BCAST is an optimized neighbor knowledge based broadcast protocol provides robust performance with less delay time (keeps two hop neighbor information and minimizes network congestion) and less traffic overhead (partial source route) in terms of PDR, latency, normalized routing and MAC load, packet drop probability and data throughput. On-demand unicast DSR protocol suffers more from scalability issue as the number of data senders and data receivers increase. For scenarios with *N* senders and *M* receivers, $N \times M$ unicast connections have to be discovered and maintained by the underlying unicast routing algorithm. This introduces substantial overhead and causes high network load. Unicast DSR is also more sensitive to data send rate. The simulation result shows that the broadcast protocol BCAST work very well in most scenarios and are more robust even with high traffic environments.

AUTHORS PROFILE

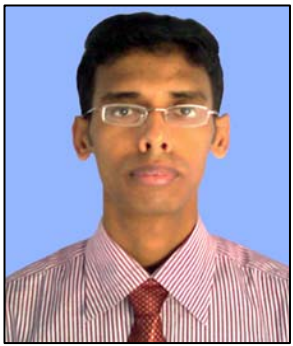

**Sumon Kumar Debnath** obtained his M.Sc degree in Information & Communication Engineering from Rajshahi University, Bangladesh during the year 2009. He has joined Noakhali Science & Technology University, Bangladesh and working as a lecturer in the dept. of Computer Science and Telecommunication Engineering. He is also working as an Assistant Proctor in that university. His research interests include Mobile Ad-hoc networks, Vehicular Ad-hoc networks, Sensor Network, MIMO, OFDM, and MCCDMA.

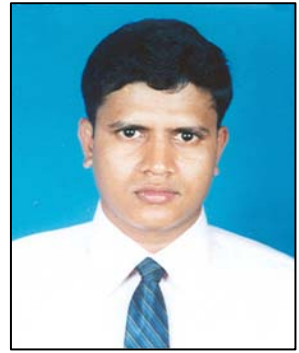

**Foez Ahmed** joined as a Lecturer in the Department of Electronics and Communication Engineering, Northern University Bangladesh (NUB), Dhaka, Bangladesh in the year of 2008. At present he is on leave from Northern University and working as a Lecturer with the Dept of Networks and Communication Engineering, College of Computer Science, King Khalid University, Kingdom of Saudi Arabia. He did his under graduation and post graduation in Information and Communication Engineering in 2007 and 2009 respectively from Rajshahi University, Bangladesh. He has received various Awards and Scholarships for the under graduation and post graduation results. His research interests include Mobile Ad-hoc Networks and Routings, Cognitive Radio Networks, Cooperative Communications, Sensor Networks and Sparse Signal Processing in Wireless Communication.

**Nayeema Islam** was born 1976 in Bangladesh. She received her M.S. in Telecommunication Engineering from Asian Institute of Technology, Thailand in 2004. Also she received her M.Sc. and B.Sc. in Computer Science and Technology from Rajshahi University, Bangladesh in 1998 and 1997 respectively. She is presently working as Assistant Professor in the department of Information and Communication Engineering, University of Rajshahi since 2004. Her fields of interest include Telecommunications, computer networking, mobile ad-hoc networks and QoS routing.